\documentclass[twocolumn,prb,aps,amsmath,amssymb,floatfix,showpacs,longbibliography,notitlepage,superscriptaddress]{revtex4-1}

\usepackage[usenames,dvipsnames]{xcolor}
\usepackage{graphicx}
\usepackage{dcolumn}
\usepackage{bm}
\usepackage{subfigure}
\usepackage[colorlinks=true,linkcolor=blue,citecolor=blue,urlcolor=blue]{hyperref}
\usepackage[normalem]{ulem}
\usepackage[utf8]{inputenc}
\DeclareUnicodeCharacter{2212}{-}

\graphicspath{{./figs/}}
\begin{document}

\newcommand{\bo}{\boldsymbol}
\newcommand{\boq}{\mathbf{q}}
\newcommand{\bok}{\mathbf{k}}
\newcommand{\bor}{\mathbf{r}}
\newcommand{\boG}{\mathbf{G}}
\newcommand{\boR}{\mathbf{R}}
\newcommand\2{$_2$}

\newcommand\theosmarvel{Theory and Simulation of Materials (THEOS), and National Centre for Computational Design and Discovery of Novel Materials (MARVEL), \'Ecole Polytechnique F\'ed\'erale de Lausanne, CH-1015 Lausanne, Switzerland}
\newcommand\geneva{Department of Quantum Matter Physics, University of Geneva, CH-1211 Geneva, Switzerland}

\title{Valley-Engineering Mobilities in 2D Materials}

\author{Thibault Sohier}
\affiliation{\theosmarvel}
\author{Marco Gibertini}
\affiliation{\geneva}
\affiliation{\theosmarvel}
\author{Davide Campi}
\affiliation{\theosmarvel}
\author{Giovanni Pizzi}
\affiliation{\theosmarvel}
\author{Nicola Marzari}
\affiliation{\theosmarvel}

\date{\today}


\begin{abstract}
Two-dimensional materials are emerging as a promising platform for ultrathin channels in field-effect transistors. To this aim, novel high-mobility semiconductors need to be found or engineered. While extrinsic mechanisms can in general be minimized by improving fabrication processes, the suppression of intrinsic scattering (driven e.g. by electron-phonon interactions) requires to modify the electronic or vibrational properties of the material. 
Since intervalley scattering critically affects mobilities, a powerful approach to enhance transport performance relies on engineering the valley structure. We show here the power of this strategy using uniaxial strain to lift degeneracies and suppress scattering into entire valleys, dramatically improving performance. This is shown in detail for arsenene, where a $2\%$ strain stops scattering into 4 of the 6 valleys, and leads to a $600\%$ increase in mobility. The mechanism is general and can be applied to many  other materials, including in particular the isostructural antimonene and blue phosphorene.
\end{abstract}

\maketitle

\begin{figure}[ht]
\includegraphics[width=0.32\textwidth]{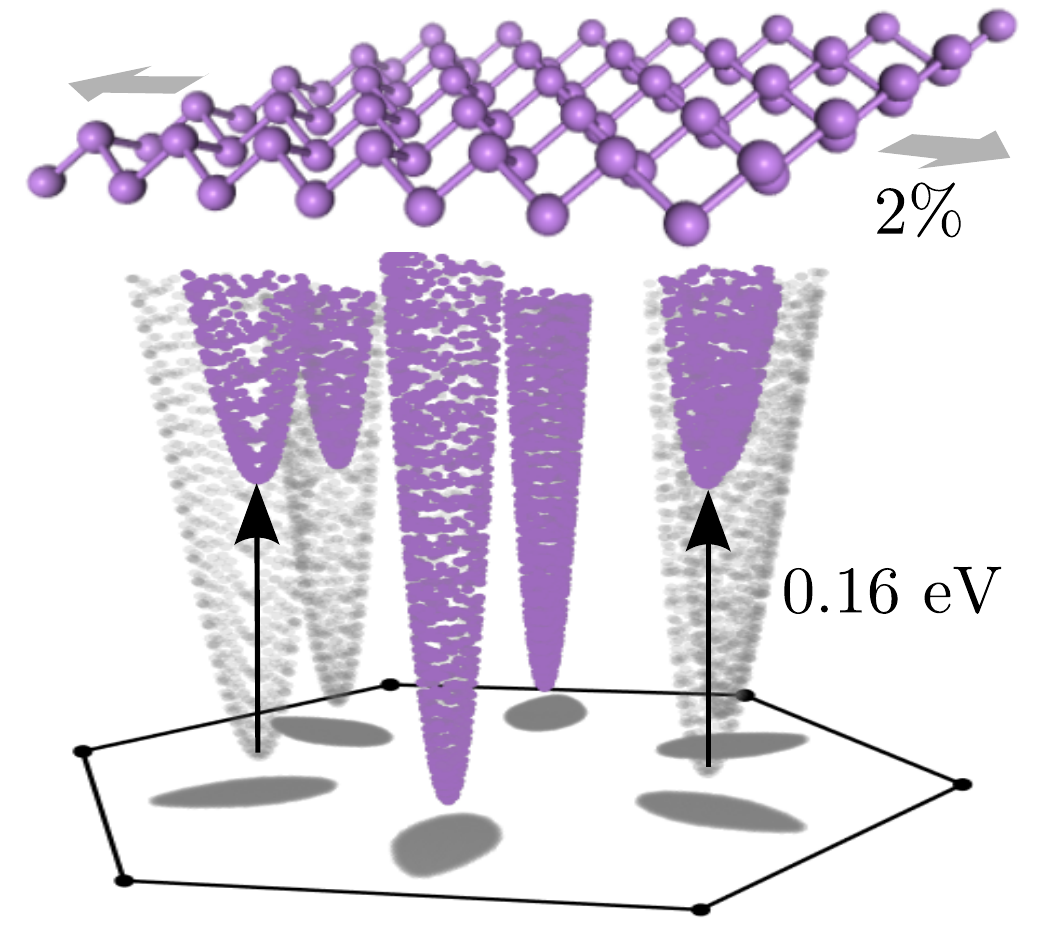}
\caption{Graphical Abstract}
\end{figure}

In the last decade, two-dimensional (2D) materials have shown a wealth of novel physical phenomena, and manifold possible technological applications. 
Their ultimate thinness is particularly suitable for applications in conventional and flexible electronics \cite{Wang2012,Fiori2014,Chhowalla2016,Wang2014,Das2014,Akinwande2014}, where 2D semiconductors could be exploited as channels in field-effect transistors. An essential ingredient to boost performance of both logical and radio-frequency devices is a large carrier mobility. Up to now, the largest room-temperature mobility in 2D materials has been reported in graphene\cite{Chen2008,Dean2010,Efetov2010,Mayorov2011,Wang2013Nov}, with optimal results in finite-frequency applications\cite{Schwierz2010,Grigorenko2012,Tassin2013}.  Nonetheless, the absence of an energy gap prevents its use in logical devices. Other well-known 2D materials, such as transition metal dichalcogenides (TMD), show large on-off ratios\cite{Yoon2011,Radisavljevic2011} but display relatively poor carrier mobilities\cite{Radisavljevic2011,Allein2014,Jo2014}. 

There is thus a pressing need to find 2D semiconductors with a finite band gap and, at the same time, high conductivity.
To design or discover such materials one must minimize 
the scattering processes degrading electrical transport. 
Considering that the quality of samples can typically be improved systematically to 
suppress external scattering sources, such as impurities, design strategies
should focus on maximizing intrinsic performance. 
In this respect, particular attention can be devoted to electron-phonon interactions, as phonons play a dominant role in limiting the intrinsic mobility of a material, especially at room temperature. In particular, the presence of multiple valleys within an energy range compatible with phonon frequencies leads to plentiful phonon-assisted intervalley scattering, that has detrimental effects on mobility. Indeed, even if this is not the only scattering mechanism present, we have recently shown~\cite{Sohier2018} that intervalley electron-phonon coupling is large in well-known 2D semiconductors, to the point that there is a strong correlation between the number of valleys and the intrinsic mobility. This suggests a potential strategy to enhance electronic performance by engineering the valley structure to suppress intervalley electron-phonon scattering. 

To achieve this goal, the extreme sensitivity of 2D materials to external manipulations can be beneficial, as it allows to easily tune electronic properties. As an example, 2D materials can be easily stretched, and it has been shown that strain plays an important role in determining  the electronic and optical properties of 2D materials\cite{Roldan2015,Mohiuddin2009}. For instance, strain fields open a gap in graphene on BN \cite{San-Jose2014,Jung2015}, change the gap from direct to indirect in antimonene \cite{Kripalani2018} or TMDs\cite{Steinhoff2015}, allow the manipulation of spin and valley transport  \cite{Lee2017a,Ma2016}, and reduce the band gap of TMDs\cite{Shen2016}.     

Here we explore the strong dependence of mobility on valley structure, and showcase that using strain is a very viable way to engineer the valley structure of 2D materials, leading to remarkable improvements in conductivity. We examine first arsenene\cite{Zhang2015,Shah2018,BeladiMousavi2019} as a prototypical example, owing to its six-fold multivalley conduction bands, and show that a relatively small uniaxial strain is sufficient to break the six-fold symmetry and push four of these arsenene valleys high enough in energy to effectively suppress all the corresponding intervalley scattering processes, in turn greatly enhancing electron mobility.   This effect is more general than in MoS\2~\cite{Ge2014} (where it is very specific to the different character of inequivalent valleys and their behaviour under biaxial strain) and much larger than in conventional 3D multivalley semiconductors\cite{Sun2007,Takagi2008,Yu2008,Chu2009,Xu2012,Vogelsang1992,Vogelsang1993,Takagi1996,Rahman2005,Uchida2005}, such as silicon and germanium, with an enhancement factor that will be shown to be, for arsenene, of the order of 600\%. 

We stress that what is suggested here is at variance with recent theoretical studies that have proposed strain as a way to enhance mobility in various 2D materials\cite{Yu2016,Aierken2016,Shao2017,Xu2017,Phuc2018,Fang2019,Liu2015,Priydarshi2018,Xie2019}.
These studies neglect {\it inter}valley scattering and consider the effect of strain on effective masses and, partially, on the {\it intra}valley scattering. Moreover, most rely on an acoustic-deformation-potential approximation\cite{Takagi1994} to compute mobilities, which seldom provides quantitative realistic predictions. 
Here, instead, we adopt  an accurate and systematic approach to compute phonon-limited mobilities \cite{Giustino2017} that combines Boltzmann transport~\cite{Ziman,Grimvall} with band and momentum-resolved scattering amplitudes obtained within density-functional perturbation theory~\cite{Baroni} (DFPT) in a framework that  accounts for reduced dimensionality and field-effect doping\cite{Sohier2017,Sohier2018}. The strength of the approach is to provide a full and detailed picture of electron-phonon scattering by explicitly computing all  possible phonon-mediated transitions, with a predictive power that can provide quantitative agreement with experiments\cite{Park2014,Sohier2014a}. 

While we perform here an in depth study of arsenene (a monolayer of arsenic atoms forming a buckled hexagonal honeycomb structure) as a reference system, similar results can be expected for other group-V buckled monolayers,  or other multivalley 2D materials. In the Supporting Information it is shown that the same features are also present in antimonene~\cite{Ji2016} and blue phosphorene~\cite{Zhang2018}.  
In addition, a search on our database of 2D materials ~\cite{Mounet2018} revealed that up to one third of semiconductors have degenerate multivalley electronic structures at the conduction or valence band edges, and could thus be valley-engineered. 
Also, although strain is very effective in modifying the valley structure of  2D materials, other mechanisms could be put forward to achieve the same scope (e.g. via stacking, substrate choice, functionalization). We thus stress that what we propose is more general than the specific compound (arsenene) and mechanism (strain) that we consider to exemplify the strategy of valley-engineering the mobility of 2D materials.

\begin{figure*}[t]
\includegraphics[width=0.65\textwidth]{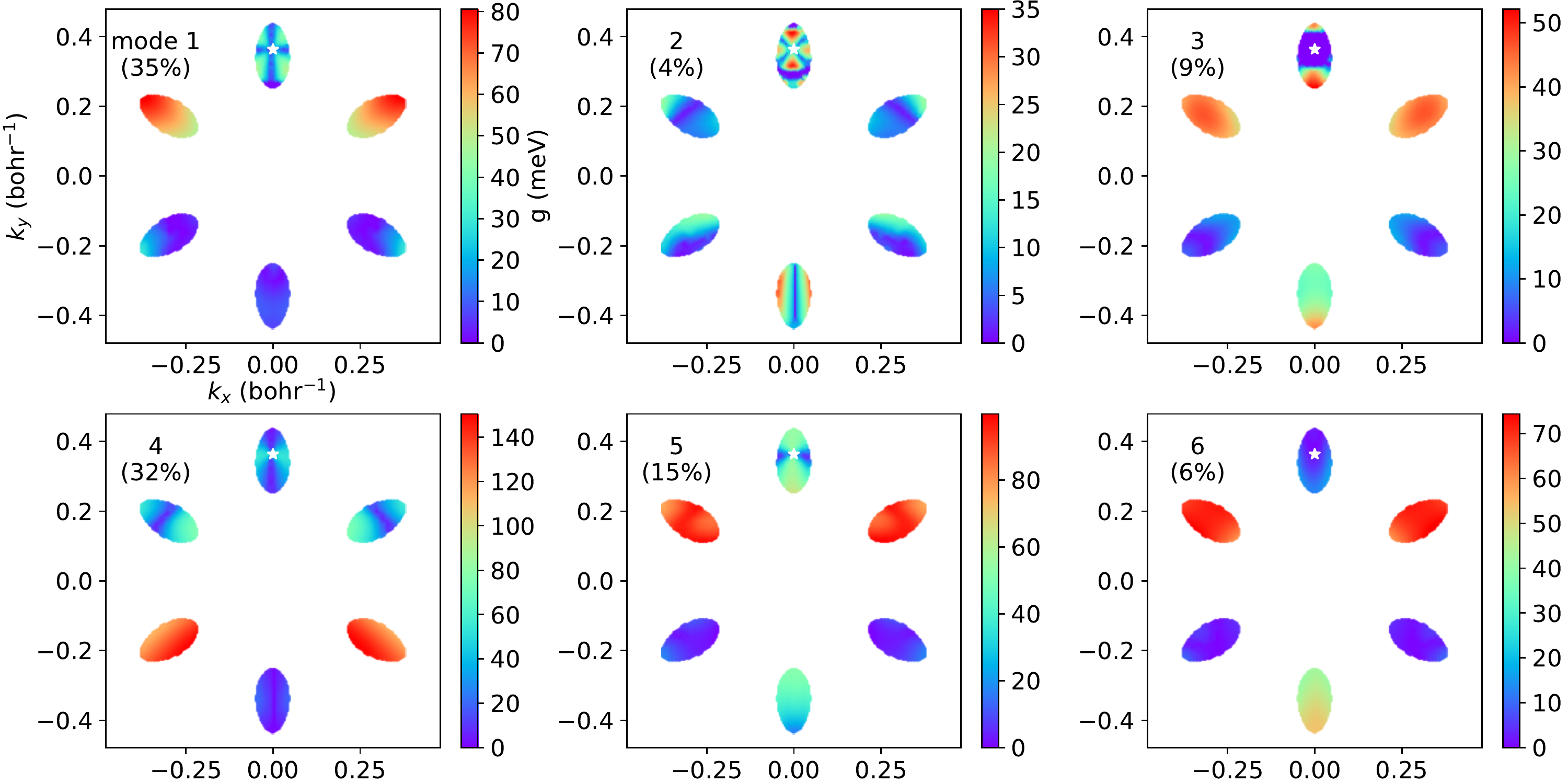}
\includegraphics[width=0.34\textwidth]{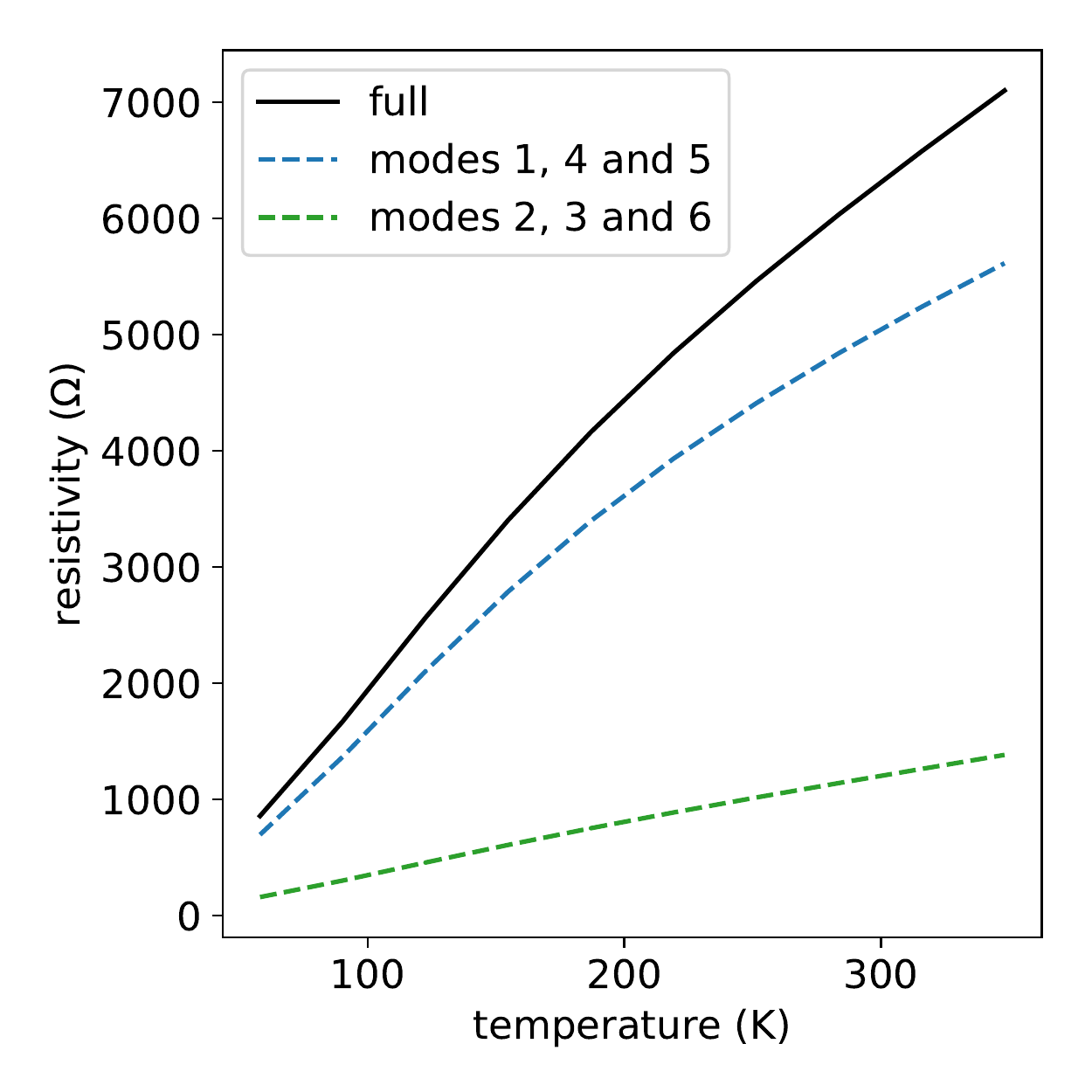}
\caption{ {\bf Left:} Interpolated electron-phonon couplings
for $n-$doped arsenene. The initial state $\bok$ considered here is indicated by a white star; the other points are the
possible final states, where the color of the point indicates the strength of the electron-phonon coupling matrix element (note the different color scales in each panel). The index of the phonon mode indicated at the top of each panel refers to a purely energetic ordering of the phonon modes associated with each transition.
Below that we indicate the contribution of each mode to the resistivity, evaluated by solving the Boltzmann transport equation at 300 K for each mode independently. While not strictly valid, as the contributions are not additive, this decomposition gives a reasonable estimate of the relative importance of each mode, highlighting that most of the resistivity is due to intervalley scattering from modes 1, 4 and 5. {\bf Right:} Plot of the resistivity as a function of temperature, as well as the contribution from the three intervalley modes (1, 4, 5) and the other modes (2, 3, 6).
The room-temperature mobility that we compute is $60$~cm$^{2}$V$^{-1}$s$^{-1}$.}
\label{fig:res_contrib_As}
\end{figure*}

In the following, the zig-zag (armchair) direction is aligned with the $\bo{\hat x}$ ($\bo{\hat y}$) axis.
For transport purposes, the conduction band of arsenene can be considered as consisting of six equivalent anisotropic valleys along the $\Gamma-$M lines, as shown in Fig.~\ref{fig:res_contrib_As}. Indeed, the next low-energy conduction band minima 
occur at least $0.27$ eV above the bottom of those valleys. 
We consider all electronic states up to $0.27$~eV above the bottom of the conduction band, and we have verified that this is enough to include all relevant states for transport for the doping levels and temperatures (up to 350 K) considered in this work. 

Despite the anisotropy of each individual valley manifest in Fig.~\ref{fig:res_contrib_As}, the total mobility 
of arsenene is isotropic because of the hexagonal symmetry of the system. This has been shown both analytically\cite{Pizzi2016a}
and numerically\cite{Sohier2018}, and is expected from general symmetry considerations\cite{Nye}. 
For completeness and clarity, we summarize here a general derivation that will be useful to elucidate
transport properties under the application of strain.
The total conductivity is a sum of contributions from each valley:
\begin{align}
\sigma &= 2e^2 \sum_p  \int_{\Omega_p} \frac{d\bok}{(2\pi)^2} 
(\bo{v}_p(\bok) \cdot \bo{u}_{E})^2 \tau_p(\bok) 
\frac{\partial f^0}{\partial \varepsilon_{\bok}},
\end{align}
where $p=1,\ldots,6$ is the index running over the six valleys 
(we choose the reference valley $p=1$ as the one in the $+\bo{\hat y}$ direction, so that its principal directions simply coincide with the Cartesian axes), $e$ is the electron charge,
$\bo{v}_p(\bok)$ are the band velocities, $\tau_p(\bok)$ are the $\bok-$dependent
scattering times, $f^0$ is the equilibrium Fermi--Dirac distribution,
$\bo{u}_{E}$ is a unit vector in the direction of the applied electric
field, and the integral is over the area $\Omega_p$ around each valley
$p$ where $\frac{\partial f^0}{\partial \varepsilon_{\bok}}$ is significantly
different from zero.
The contributions from each valley are related to each other by symmetry.
Indeed, all valleys can be obtained by rotating the reference valley 
by angles $ \theta_p \in [\pi/3, 2\pi/3,\pi , 4\pi/3, 5\pi/3]$. 
Eigenenergies and scattering times are invariant under these operations, and only the projection of the velocity on the electric field direction
$\bo{v}(\bok) \cdot \bo{u}_{E}$ changes.
We can thus obtain the conductivity equivalently by carrying out the
integral in the reference valley ($p = 1$) and rotating the electric field 
to add up the contribution of all valleys:
\begin{align}
\sigma &= 2e^2   \int_{\Omega_1} \frac{d\bok}{(2\pi)^2} 
\left[ \sum_p (\bo{v}_p(\bok) \cdot \bo{u}_{p})^2\right] 
\tau_1(\bok) \frac{\partial f^0}{\partial \varepsilon_{\bok}},
\end{align}
where $\bo{u}_{p} = \cos(\theta_E-\theta_p) \bo{\hat x} + \sin(\theta_E-\theta_p) \bo{\hat y}$, 
with $\theta_E$ giving the direction of the applied electric field. 

By expanding the squared scalar product and noting that the 
corresponding cross product vanishes because of the mirror symmetry with respect to the $\Gamma-$M lines, we get:
\begin{align}\label{eq:cond}
\sigma &= \sigma_h \sum_p \cos^2(\theta_E-\theta_p) +\sigma_l \sum_p
\sin^2(\theta_E-\theta_p), \text{ with} \\ 
\sigma_{h,l} &\equiv \sigma_{x,y} = 2e^2   \int_{\Omega_1} \  \frac{d\bok}{(2\pi)^2}\
    v^2_{1 x,y}(\bok) \tau_1(\bok) \frac{\partial f^0}{\partial \varepsilon_{\bok}},\label{eq:sigma-h-l}
\end{align}
where the conductivity in the $\bo{\hat x}$ ($\bo{\hat y}$) direction for the 
reference valley will be denoted with the $h$ (or $l$) subscript because it corresponds to the high (or low) velocity and thus indicates the direction with the high (or low) conductivity.
Summing up for $\theta_p \in [0, \pi/3, 2\pi/3,\pi , 4\pi/3, 5\pi/3]$ 
yields:
\begin{align}
\sigma &= 3 (\sigma_h+\sigma_l)\label{eq:cond-unstrained-final}
\end{align}
independently of the direction $\theta_E$ of the field, proving that, even though the conductivity of a single valley is anisotropic ($\sigma_h\neq\sigma_l$), the overall conductivity of the material is indeed isotropic~\cite{Nye}.

A detailed analysis of electron-phonon scattering in arsenene 
can help us reveal the strength of intervalley scattering and 
which intra- and intervalley couplings play the largest role in inhibiting transport performance. 
In particular, we show in Fig.~\ref{fig:res_contrib_As} the
electron-phonon couplings (EPCs) for each phonon mode and their contribution to transport. We can see that $\approx$80\% of the resistivity comes from scattering with
modes associated with strong intervalley EPCs, corresponding to phonon
momenta $\boq \approx K$ and $\boq \approx \frac{\Gamma-M}{2}$. 
Considering an initial state in the reference valley, these modes correspond
to transitions to valleys rotated by $\theta = \pm \pi/3, \pm 2\pi/3$. 

It would thus be beneficial to suppress scattering to these valleys to enhance the electronic transport performance. 
Valleys in different directions have the same energy because 
of the 3-fold and inversion symmetries of the crystal. 
Breaking these symmetries removes the degeneracy and allows for the emergence of an 
energy difference between valleys. 
If this energy difference is large enough, 
valleys that are shifted up in energy become out of reach, i.e., electrons cannot be scattered into them via a phonon-mediated event. In this respect, it is important to keep in mind that, under the assumption of constant charge (see below), the remaining valleys contain more carriers, so that the Fermi energy is also shifted up in energy with respect to the bottom of the conduction band, which is fully accounted for in our calculations and discussed in details later.
As a consequence, in order to fully suppress intervalley scattering, the shift in energy of the valleys should be large enough so that the bottom of the valleys is higher than the (new) Fermi level by at least the largest phonon energy (relatively low in arsenene, $< 0.03$~eV for the relevant phonons involved in intervalley scattering events) plus the thermal energy (also $< 0.03$~eV at 300 K).

We consider a strain along the $\bo{\hat x}$ (zig-zag) direction that pushes all the valleys that are not on the $\bo{\hat y}$ axis (4 out of 6, see also Fig.~\ref{fig:taus}) high in energy with respect to the band edge, as opposed to a strain along the $\bo{\hat y}$ (armchair) direction that would instead shift only two valleys to higher energy\footnote{According to Ref.~\onlinecite{Pizzi2016a} the deformation potential of the conduction valleys is positive in the longitudinal direction and negative in the transverse direction, so that with a strain along the $\bo{\hat x}$ direction two valleys (the ones with longitudinal direction along $\bo{\hat y}$) are pushed down and the other four are pushed up, while the opposite is true (two up and four down) for strain along the $\bo{\hat y}$ direction.}. In particular, we find that a $2\%$ uniaxial strain\footnote{in practice, depending on the experimental setup, such an uniaxial stretch might be accompanied by a compression in the perpendicular direction. This is not included in our simulations, but we checked that the effects of a full relaxation in the perpendicular direction are minimal.} induces a largely sufficient energy shift of $\approx 0.16$ eV. 
Such a strain is readily within the possibilities of experimental realization~\cite{Mohiuddin2009,Huang2009} and can be indirectly monitored through its effects on arsenene's Raman spectrum, in particular through the splitting of the E$_{g}$ mode (see predicted Raman spectra under strain and their polarisation dependence in Supporting Information F).
In antimonene and blue phosphorene, the same strain shifts the same valleys by $\approx 0.10$ and $0.22$ eV, respectively (also see Supporting Information D and E). Note that according to  Ref.~\onlinecite{Pizzi2016a} (supplementary figure 5), one can roughly assume a linear variation of the shifts with strain. 

We simulate equilibrium and strained arsenene in the appropriate two-dimensional
framework\cite{Sohier2017} by truncating the long-range Coulomb potential in the non-periodic direction. This ensures that no spurious interactions exist between the periodic images, yielding in particular the correct 2D screening of the electron-phonon coupling. We also use a symmetric double-gate geometry\cite{Sohier2017} to induce a charge density of $n=5/3 \times  10^{13}$ cm$^{-2}$. 
The Brillouin zone is sampled with a $32 \times 32 \times 1$ Monkhorst-Pack grid and we use the SSSP Accuracy (version 0.7) pseudopotential library\cite{Prandini2018} with the associated cutoffs.
We note that the density of states (DOS) at the bottom of the conduction band is approximately three times smaller for strained arsenene, since two thirds of the valleys have been shifted up. Thus, the two remaining low-lying valleys fill up three times faster as a function of the doping charge. The same density of free carriers is considered in both the strained and unstrained cases, under the assumption that the gate capacitance is mainly controlled by geometric effects and that the reference potential of the material is almost unaffected, so that the density is completely controlled by the gate potential and thus stays constant under strain. The reference value of density ($n=5/3\times 10^{13}$ cm$^{-2}$) 
is then chosen so that the shifted valleys in the strained case are not occupied, 
while the valleys in the pristine case are still significantly filled. 

\begin{figure*}[t]
\includegraphics[width=0.90\textwidth]{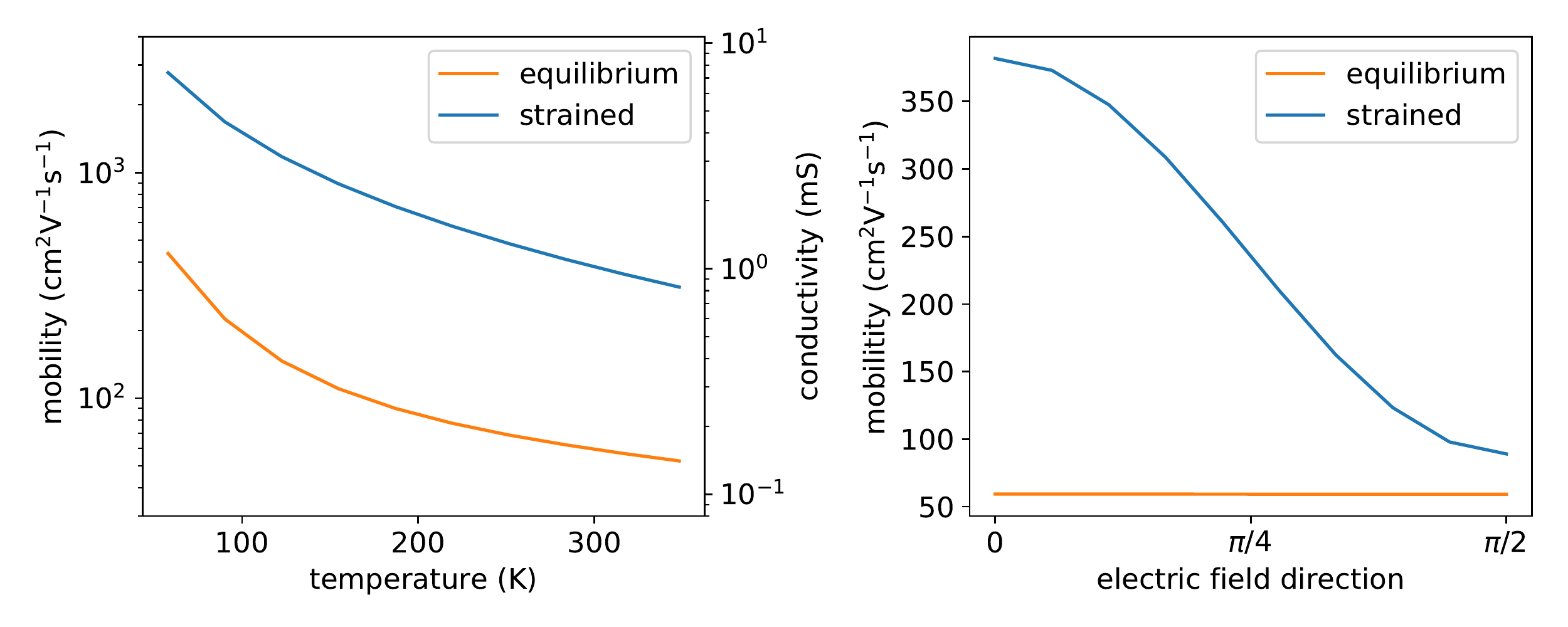}
\caption{\textbf{Left:} Conductivity and mobility along the $\bo{\hat x}$ direction as a function of temperature for equilibrium and strained ($2\%$ in the $\bo{\hat x}$ direction) arsenene. 
\textbf{Right:} Dependency of the room-temperature mobility on the direction of transport. $\theta_E=0$ indicates a field aligned along the $\bo{\hat x}$ direction. }
\label{fig:mobility}
\end{figure*}

Mobilities are computed with the method described in Ref.~\onlinecite{Sohier2018}, which is based on direct calculation of phonons and EPCs within DFPT in the
aforementioned framework including the proper boundary conditions and
explicit gate-induced doping. 
The Boltzmann transport equation is solved
numerically with an exact integration method and different levels of momentum  sampling, from a relatively coarse $32 \times 32 \times 1$ $\bok$-point grid used for DFPT calculations to a much finer $177 \times 177 \times 1$ $\bok$-point grid for eigenenergies and velocities. 
The transport properties that we computed for equilibrium and strained
arsenene are compared in Fig.~\ref{fig:mobility}. 
For a $2\%$ strain,  we observe a significant improvement of the transport properties for an in-plane electric field (source-drain bias) in the $\bo{\hat x}$ direction, i.e., the high-velocity direction of the two low-energy conduction valleys. Namely, the conductivity and the mobility increase by a factor $\alpha_h \approx 6.4$. This factor is the same for both quantities as we are working at constant density. However, the factor depends on the direction of the electric field as the valleys are anisotropic and the strained system does not have hexagonal symmetry. Nevertheless, even in the orthogonal low-velocity direction $\bo{\hat y}$, along which the mobility of strained arsenene reaches its minimum, we observe an increase of the mobility with respect to the unstrained case by a factor $\alpha_l \approx 1.3$.

To investigate and explain the origin of this enhancement factor, we first observe that the EPCs are essentially the same for the strained and equilibrium case, as it can be observed by comparing Fig.~\ref{fig:res_contrib_As} with an equivalent plot for the strained case (Fig.~\ref{fig:EPC} of the Supporting Information). 
Therefore, strain does not change the nature or strength of the EPCs, 
but it affects the relative energy of the states in the valleys, 
effectively turning off some intervalley scattering channels by 
moving the corresponding final states out of reach. 
From our computed scattering times taken at the Fermi level, we determine that scattering decreases by a factor $5.4$, as shown in Fig.~\ref{fig:taus},
which is consistent with the approximate mode-by-mode contributions indicated in Fig.~\ref{fig:res_contrib_As}. The suppression of intervalley scattering is thus the major contribution to the improvement of the conductivity. 

\begin{figure}[h]
\includegraphics[width=0.48\textwidth]{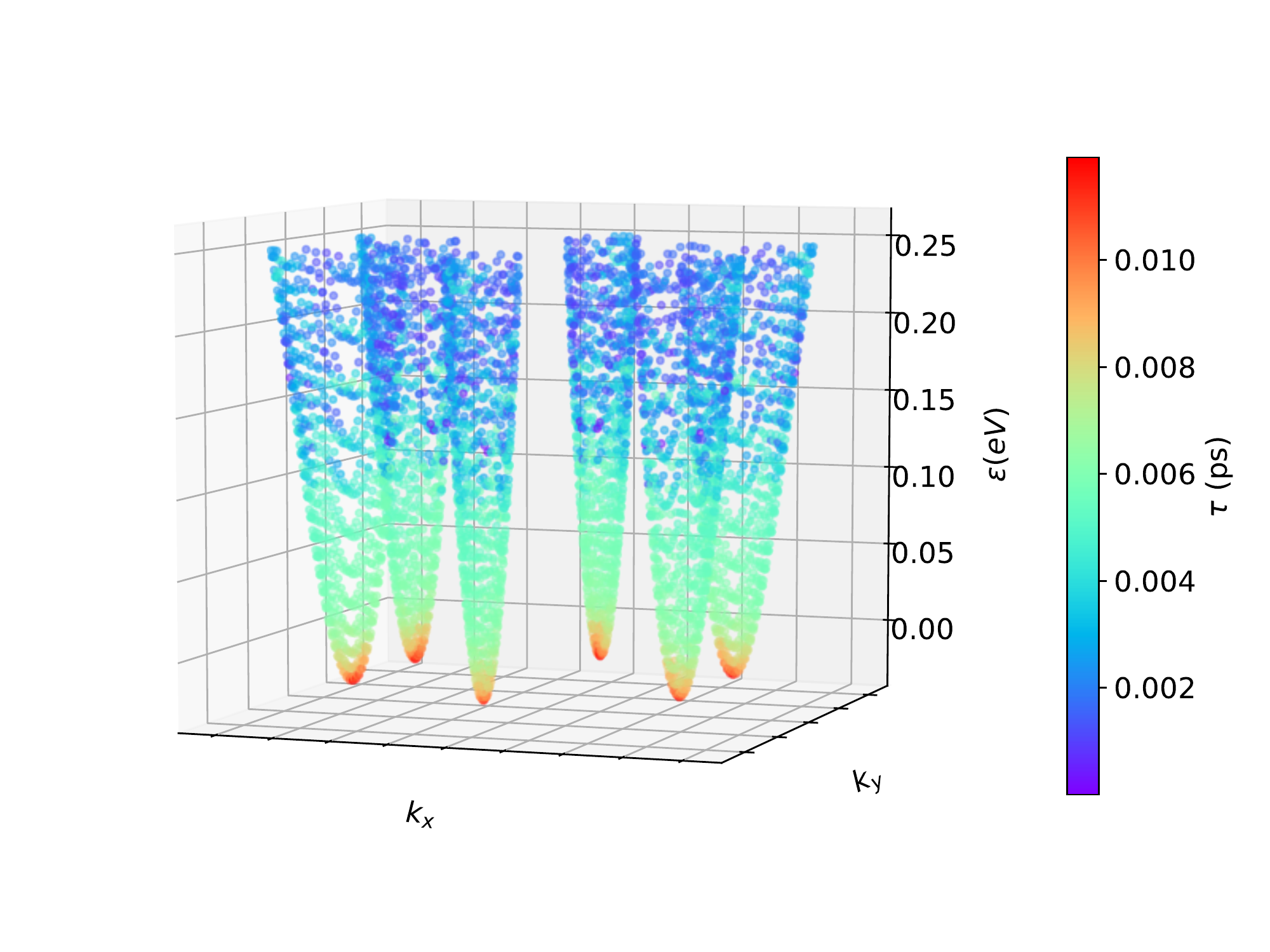}
\includegraphics[width=0.48\textwidth]{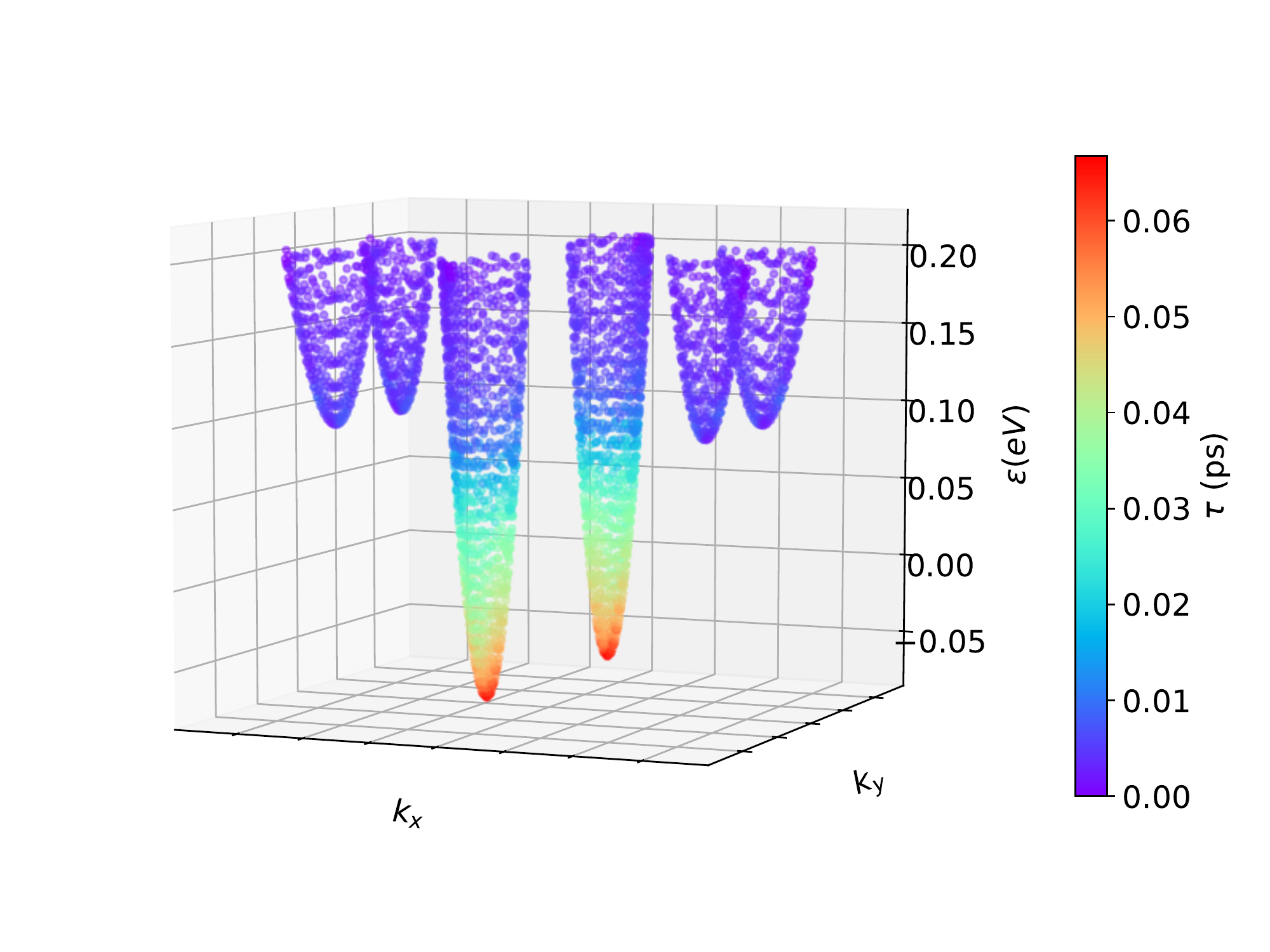}
\caption{Scattering times for equilibrium and strained arsenene. 
The ratio of the scattering times for electronic states at the Fermi level in the high-velocity direction is $\frac{\tau^{\rm{strain}}}{\tau^{\rm{equilibrium}}} 
\approx 5.4$. The zero of the energy scale refers to the Fermi level in each system.}
\label{fig:taus}
\end{figure} 

There are other contributions originating from the change of symmetry and electronic structure that must be taken into account. First, strain can affect the effective masses, although we find this effect to be negligible here (see Supporting Information C). Two other effects, however, do play a significant role, even if they happen to nearly compensate each other here. 
The first of those effects is the following: the conductivity becomes a sum of contributions from only two valleys instead of six, corresponding to a loss in density of states, and it is direction-dependent. In fact, performing the sum of Eq.~\eqref{eq:cond} over the two remaining valleys at $\theta_p \in [0,\pi]$ we get for the conductivity $\bar{\sigma}$ in the strained case:
\begin{align}
\bar{\sigma} &= 2 \bar{\sigma}_h \cos^2(\theta_E) 
+ 2 \bar{\sigma}_l  \sin^2(\theta_E),
\end{align}
The conductivity for an electric field in the high-velocity direction now is only twice the conductivity of the reference valley
($\bar{\sigma} = 2 \bar{\sigma}_h$), in contrast to $\sigma = 3(\sigma_h+\sigma_l)$ of Eq.~\eqref{eq:cond-unstrained-final} for the equilibrium case. 
Therefore, if we were to ignore the six-fold enhancement of mobility from the suppression of the intervalley scattering discussed above and assume that the contributions from a single valley did not change (i.e., $\sigma_h=\bar\sigma_h$, $\sigma_l=\bar \sigma_l$), we would actually obtain that the removal of four valleys would degrade conductivity significantly.
Single-valley conductivities in high- and low-velocity directions are numerically $\sigma_l \approx  \sigma_h /4 $; so, the conductivity in the strained case would become a factor $\frac{2 \sigma_h}{3(\sigma_h+\sigma_h/4)} \approx 0.5$ smaller in the high-velocity direction, down to a factor $\approx 0.1$ along the low-velocity direction.

\begin{figure}[h]
\includegraphics[width=0.45\textwidth]{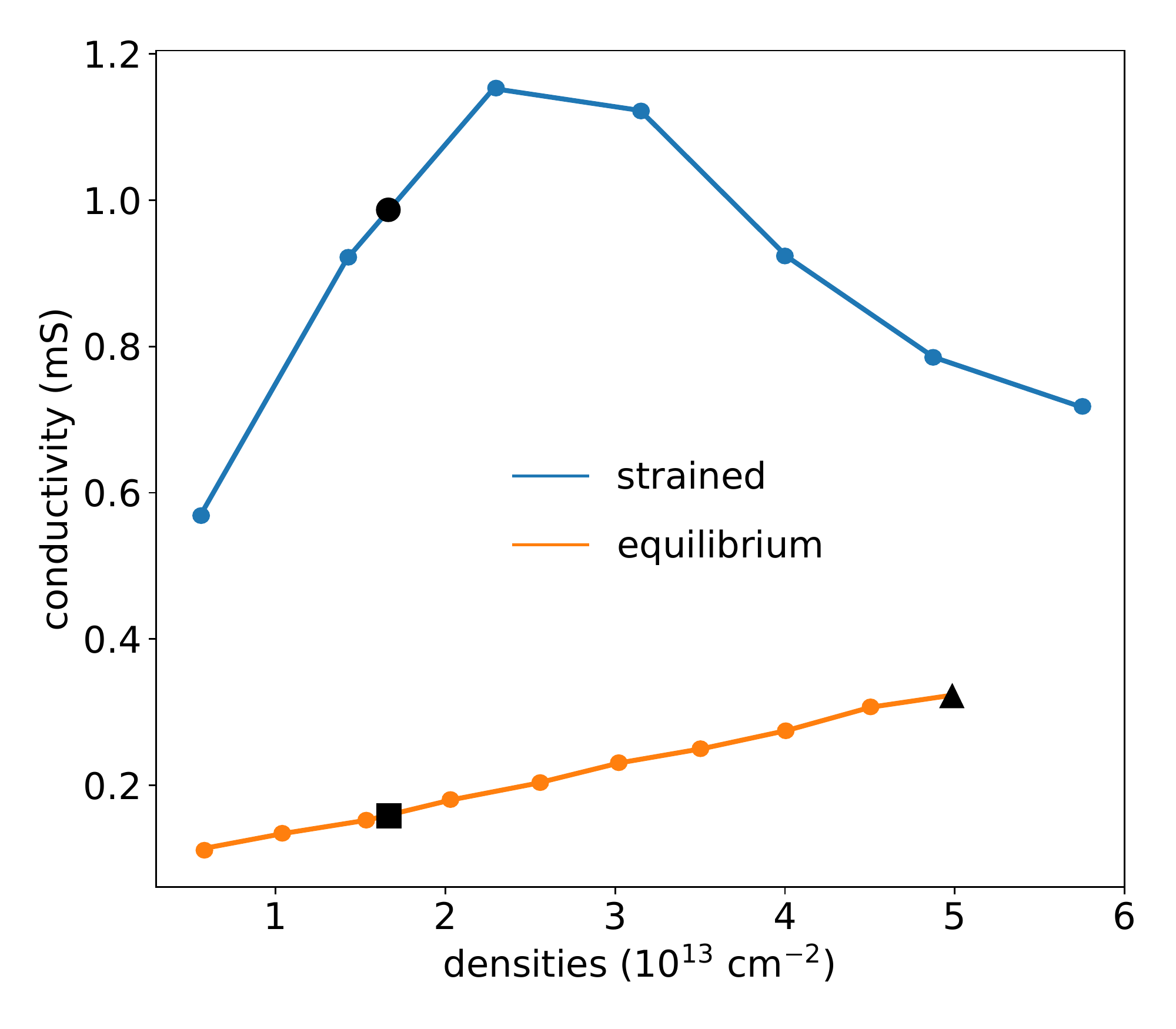}
\caption{Conductivity of strained and equilibrium arsenene as a function of carrier density. Each density corresponds to a different Fermi level (with respect to the band edge) in strained or equilibrium arsenene.
Calculations are performed in the approximation of a rigid Fermi level shift, ignoring any change in the EPC. To check that this approximation is reasonable for this material in this range, we replaced the electron-phonon coupling matrix elements in the equilibrium case with those obtained at three times higher doping (from Ref.~\onlinecite{Sohier2018}) and obtained conductivity variations below $5\%$. The square and circle correspond to the carrier density of $n= 5/3 \times 10^{13}$ cm$^{-2}$  studied in this work. The triangle corresponds to the carrier density needed in the equilibirum case for the Fermi level to reach the value of the strained case at  $n= 5/3 \times 10^{13}$ cm$^{-2}$ (circle), with respect to the band edge.}
\label{fig:Ef}
\end{figure}

However, this loss happens to be compensated by a second effect: an increase of the Fermi level within the two remaining valleys. Assuming as mentioned above that the charge density does not change under strain, the fact that only two valleys are now occupied instead of six implies that those valleys must contain three times as many carriers, which corresponds to an increase of the Fermi level with respect to the band edge.
As the overall implications on the conductivity  Eq.~\eqref{eq:sigma-h-l} are not straightforward, we study the effect of a rigid Fermi level shift on conductivity in the equilibrium and strained case. The results are reported as a function of carrier density in Fig.~\ref{fig:Ef}. Carrier density is the appropriate variable when comparing strained and equilibrium arsenene. Indeed, in an experimental measurement of transport properties as a function of strain, the carrier density would stay mostly constant, fixed by the gate. 
Note that even at the maximum doping of $5 \times 10^{13}$ cm$^{-2}$, no additional conduction-band minima need to be taken into account (band minima appear around K and $\Gamma$ for eigenenergies beyond the range considered in this work).
We highlight some points of interest. The square and circle indicate conductivities for the systems at the carrier density studied in this work.
To compare the conductivities at similar Fermi energy relative to the band edge, the density of equilibrium arsenene should be three times that of strained arsenene, corresponding to the triangle. The resulting conductivity is roughly twice larger. This also holds for strained arsenene at small enough densities, with a conductivity twice larger at $n \approx 5/3 \times 10^{13}$ cm$^{-2}$ than at $n \approx 5/9 \times 10^{13}$ cm$^{-2}$, although at densities above $2 \times 10^{13}$ cm$^{-2}$ the Fermi level would reach the four shifted valleys and the conductivity decreases. Nevertheless, staying in the doping regime of interest, the increase in Fermi level occurring in the strained case with respect to  equilibrium due to the need to fill fewer valleys with the same charge can be associated with an improvement in conductivity by a factor approximately equal to two. 
Therefore, we conclude that the increase in Fermi level and associated doubling in conductivity compensates the loss of a factor $\approx 0.5$ associated to the reduced number of valleys (at fixed scattering), leaving the major role in the six-fold improvement of the conductivity to the suppression of intervalley scattering.
Interestingly, we can also infer from Fig.~\ref{fig:Ef} that there is an optimal density, around $3\times10^{13}$ cm$^{-2}$, that maximizes the enhancement of the transport properties
for a $2\%$ strain.  Finally, we also provide an estimation of the dependency of the conductivity enhancement as a function of strain in the Supporting Information B, assuming a linear relationship between the strain and the shift of the valleys. As can be expected, the strain-dependent enhancement is not monotonic: it increases slowly at small strain and saturates at high strain.
Based on the shifts induced by strain and the computation of electron-phonon couplings in the equilibrium case, similar results are expected in antimonene and blue phosphorene (see Supporting Information).

In conclusion, we have put forward a general approach to enhance the mobility of 2D materials by engineering their valley structure. We have illustrated this approach in detail in the case of electron-doped arsenene, using a uniaxial strain field to modify the valley configuration and intervalley scattering.  
Indeed, a moderate and physically attainable strain induces a relative change in the energy of the six equivalent conduction valleys, shifting four of them out of reach for phonon-mediated scattering. 
The overall effect on the transport properties of arsenene results from a complex interplay of different mechanisms. Using accurate first-principles simulations of the electron-phonon interactions that take into account the reduced dimensionality and the field-effect doping of the system, we show that the main effect arises from the suppression of intervalley scattering, with a substantial six-fold enhancement of the electron mobility. 
Similar results can be expected also for other isoelectronic group-V buckled monolayers, such as antimonene and blue phosphorene, as well as for other multivalley 2D materials, suggesting valley-engineering as a general and viable way to design the mobility of 2D materials, with potential applications in next-generation electronics.

\acknowledgements

We acknowledge useful discussions with Alberto Morpurgo. This work was supported by the NCCR MARVEL of the Swiss National Science Foundation. G.P. acknowledges also support from the EU Centre of Excellence MaX ``Materials design at the Exascale'' (grant no.~824143). D.C.\ from the ‘EPFL Fellows’ fellowship programme co-funded by Marie Sklodowska-Curie, Horizon 2020 grant agreement no. 665667, M.G.\ from the Swiss National Science Foundation (SNSF) through the Ambizione career program (grant no.\ 174056).
Simulation time was awarded by CSCS on Piz Daint (production project s825) and by PRACE on  Marconi at Cineca, Italy (project id.\ 2016163963).

\newpage

\section{Supporting Information}

\renewcommand\theequation{S\arabic{equation}}
\renewcommand\thefigure{S\arabic{figure}}
\renewcommand\thetable{S\Roman{table}}

\setcounter{equation}{0}
\setcounter{figure}{0}
\setcounter{table}{0}
\setcounter{section}{0}

\subsection{Electron-phonon scattering in strained arsenene}

\begin{figure*}[ht]
\includegraphics[width=0.65\textwidth]{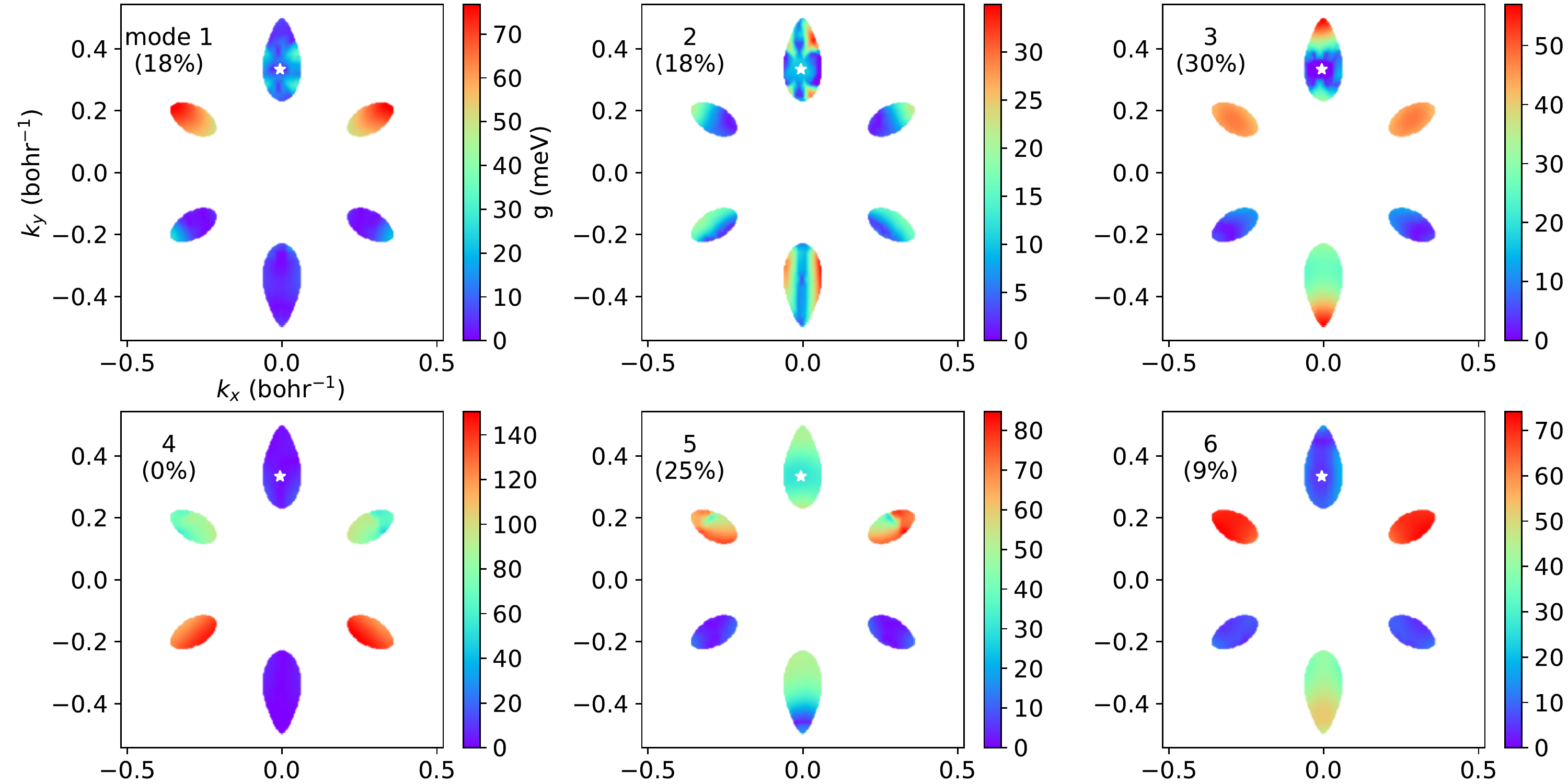}
\caption{Interpolated electron-phonon couplings in strained arsenene. The initial state $\bok$ considered here is indicated by a white star; the other points are the
possible final states, where the color of the point indicates the strength of the electron-phonon coupling matrix element (note the different color scales in each panel). The index of the phonon mode indicated at the top of each panel refers to a purely energetic ordering of the phonon modes associated with each transition.}
\label{fig:EPC}
\end{figure*}

Fig. \ref{fig:EPC} shows that the electron-phonon couplings for strained arsenene are very similar to the ones of equilibrium arsenene reported in the main text.  

\begin{figure*}[ht]
\includegraphics[width=0.55\textwidth]{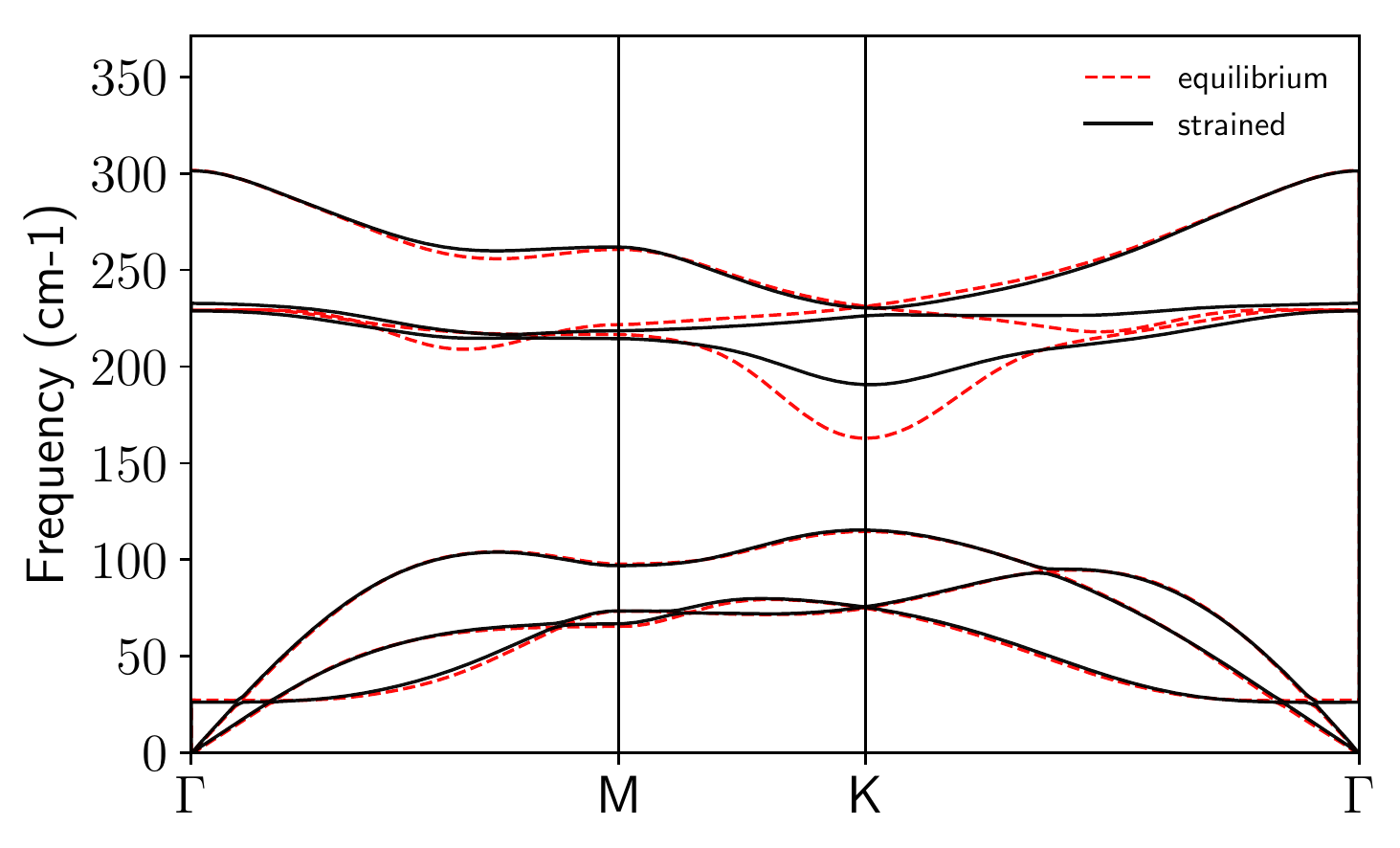}
\caption{Phonon dispersion of the equilibrium and strained case. The finite frequency of the flexural mode at $\Gamma$ is due to the presence of barriers in our simulations~\cite{Sohier2018}. The phonon softening appears in the presence of both Fermi nesting and strong EPC.}
\label{fig:phonons}
\end{figure*}

We also provide the phonon dispersion in equilibrium and  strained arsenene in Fig \ref{fig:phonons}.  

\subsection{Strain-dependent conductivity enhancement}
\begin{figure*}[ht]
\includegraphics[width=0.45\textwidth]{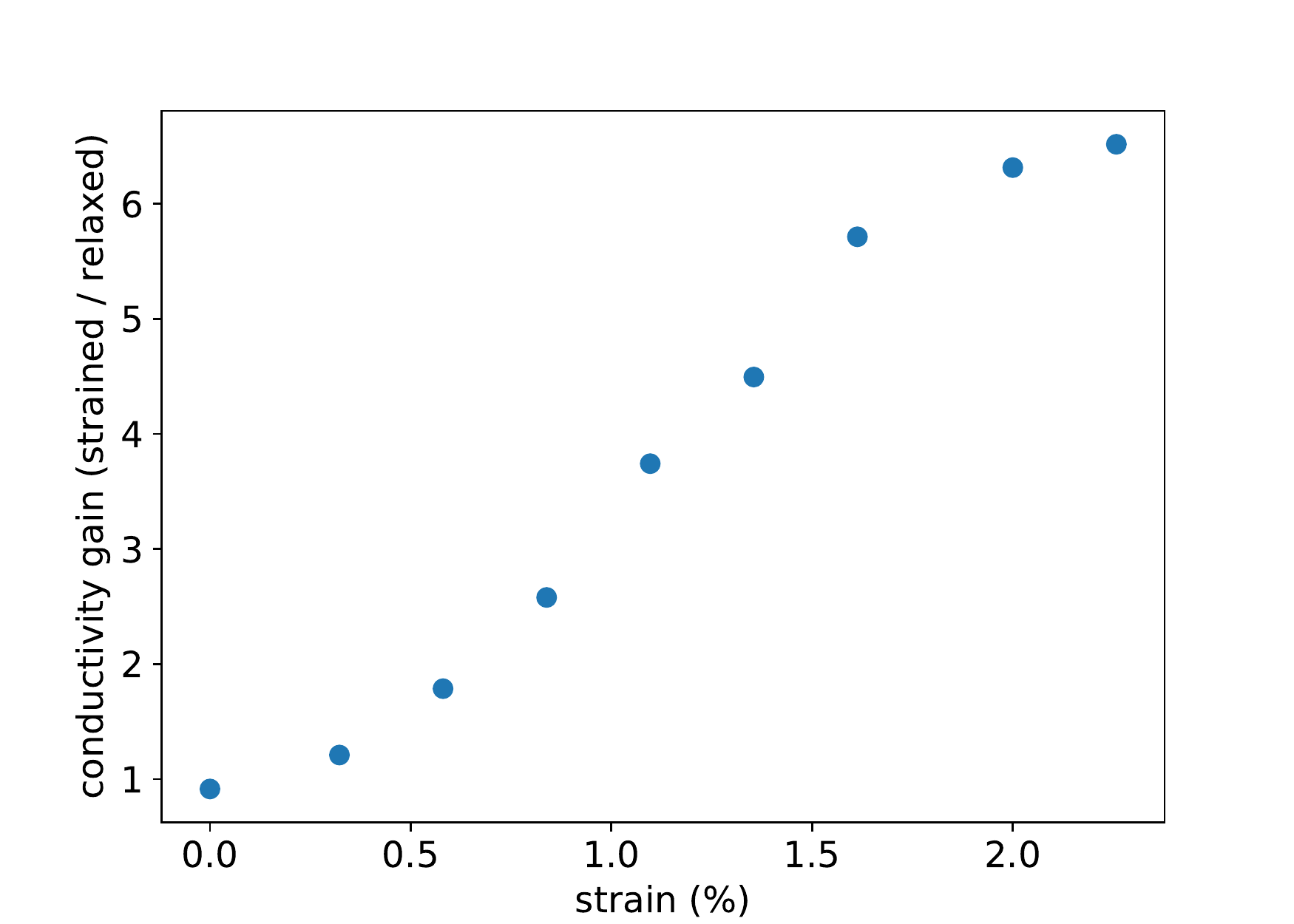}
\caption{Enhancement of the conductivity as a function of strain. For each strain, we compute the corresponding shift of the valleys assuming a linear relationship between the shift and the strain. Then, we artificially shift the valleys, and recompute the Fermi level and the conductivity using the electron-phonon coupling matrix elements of arsenene strained by $2\%$. }
\label{fig:enh_vs_strain}
\end{figure*}

In Fig. \ref{fig:enh_vs_strain}, we show an estimation of the conductivity enhancement as a function of strain. We assume a linear relationship between strain and the energy shift of the valleys. We use the electron-phonon coupling matrix elements of  arsenene strained by $2\%$. This implies that we neglect the changes of electron-phonon due to strain. This approximation has marginal effects, as  the conductivity at zero strain is well reproduced. We keep the doping carrier density constant and the Fermi level is recomputed for each strain. We see that the conductivity enhancement is not linear. It increases slowly at small strain, while the valleys are still below the Fermi level. Then we observe a steady increase of the enhancement as the valleys become less and less accessible for scattering. Finally, above $2\%$, the enhancement saturates, as the shifted valleys are already out of reach.

\subsection{Variation with strain of the effective masses}

\begin{figure*}[ht]
\includegraphics[width=0.45\textwidth]{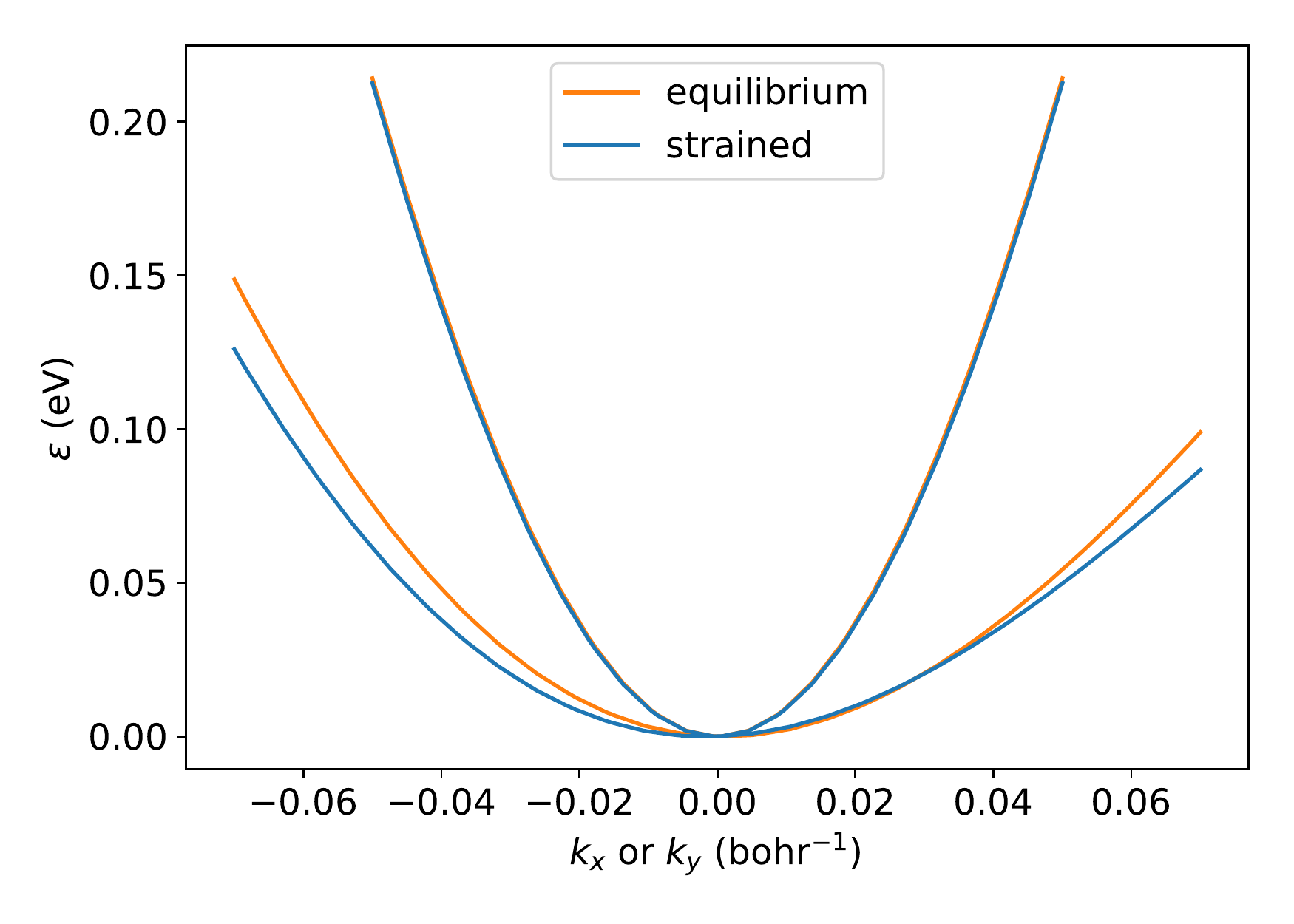}
\caption{Energy bands at equilibrium and with strain (uniaxial $2\%$) for the reference valley, situated in the $+\bo{y}$ direction, both in the high and low velocity directions ($\bo{x}$ and $\bo{y}$, respectively).  }
\label{fig:eff_m}
\end{figure*}

Fig. \ref{fig:eff_m} shows the modification of the effective masses induced by strain. The curvature decreases by $\sim 2\%$ in the high velocity direction and by $\sim 15\%$ in the low velocity direction. The effect would be squared in the conductivity, and thus induce a slight degradation of the transport properties. We are mainly interested in transport along the high velocity direction, for which we estimate a $\sim 5\%$ decrease of conductivity due to the change of effective masses. The decrease would be larger in the low velocity direction, but still negligible with respect to the other contributions discussed in the main text, which induce variations of $200\%$ or more.  

\subsection{Antimonene}
\begin{figure*}[ht]
\includegraphics[width=0.4\textwidth]{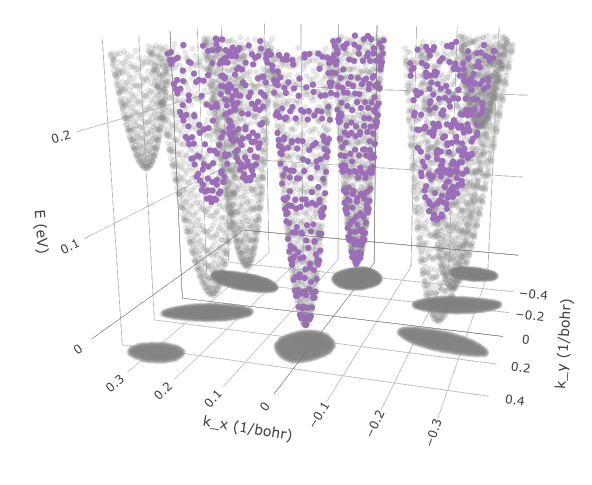}
\caption{Bands of strained (purple) and equilibrium (grey) antimonene. The strain is $2\%$ in the $x$ direction, as for arsenene. The shift in energy is more moderate $\approx 0.1$ eV. In antimonene at equilibrium, one can see that the valley at the K point is relatively low (compared to arsenene). However, this valley also shifts with strain. Here, the zero of the energy scale refers to the bottom of the lowest valley in each case. }
\label{fig:Sb_bands}
\end{figure*}

\begin{figure*}[ht]
\includegraphics[width=0.85\textwidth]{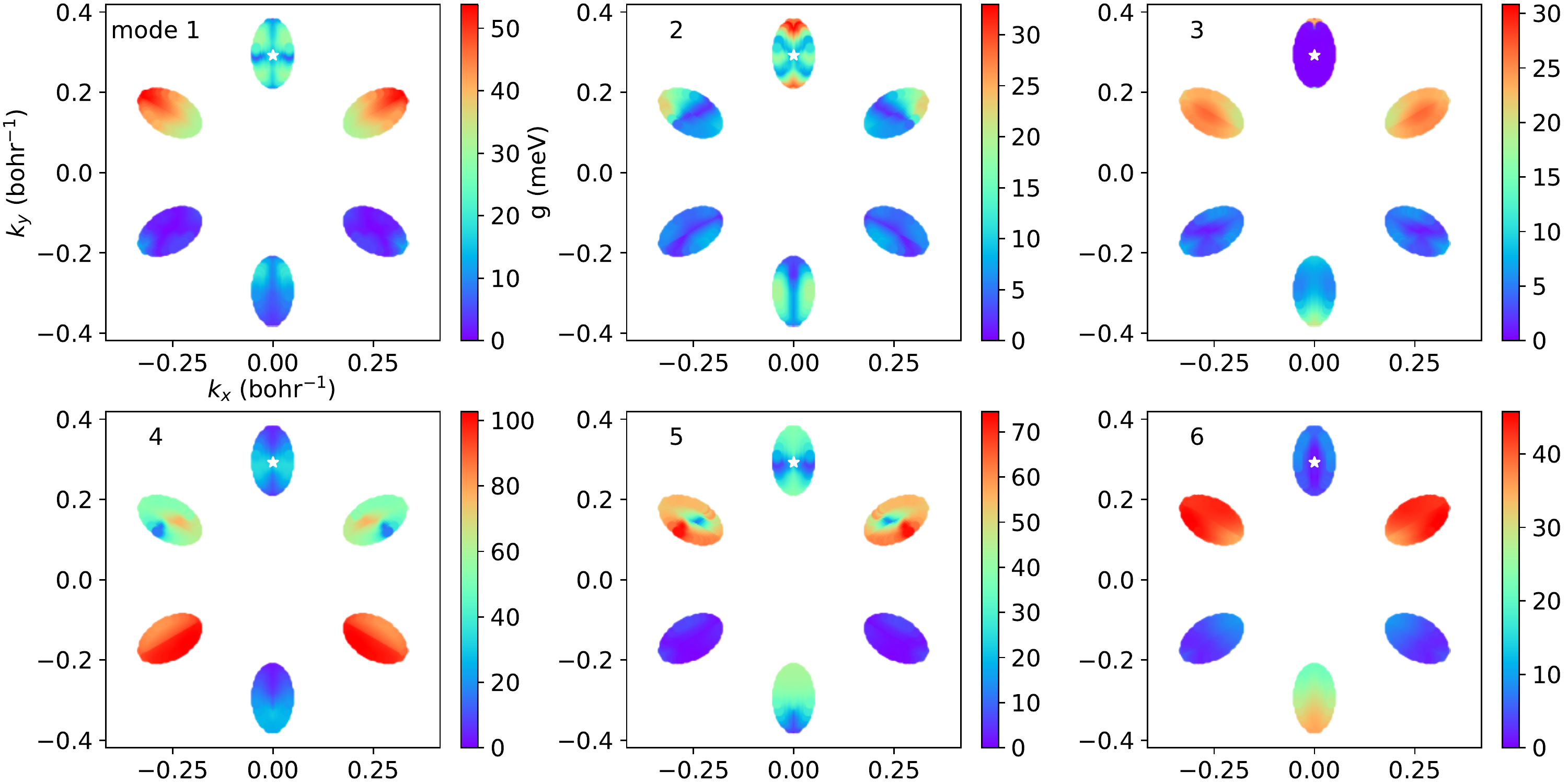}
\caption{Interpolated EPCs in antimonene at equilibrium. The initial state $\bok$ considered here is indicated by a white star; the other points are the
possible final states, where the color of the point indicates the strength of the electron-phonon coupling matrix element (note the different color scales in each panel). The index of the phonon mode indicated at the top of each panel refers to a purely energetic ordering of the phonon modes associated with each transition. }
\label{fig:Sb_EPC}
\end{figure*}

Fig. \ref{fig:Sb_bands} shows the effect of strain on the bands of antimonene. The shift in energy due to strain is smaller but still significant and enough to make the shifted valleys out of reach for phonon-mediated processes. Fig. \ref{fig:Sb_EPC} shows the electron-phonon couplings in antimonene at equilibrium. The transport properties computed from these results are very similar to those of arsenene, with a room-temperature mobility of 62 cm$^2$V$^{-1}$s$^{-1}$ for the same doping of $n = 5/3 \times 10^{13}$ cm$^{-2}$. Thus, we can expect a slightly smaller but similar enhancement of mobility in strained antimonene.

\subsection{Blue Phosphorene}
\begin{figure*}[ht]
\includegraphics[width=0.4\textwidth]{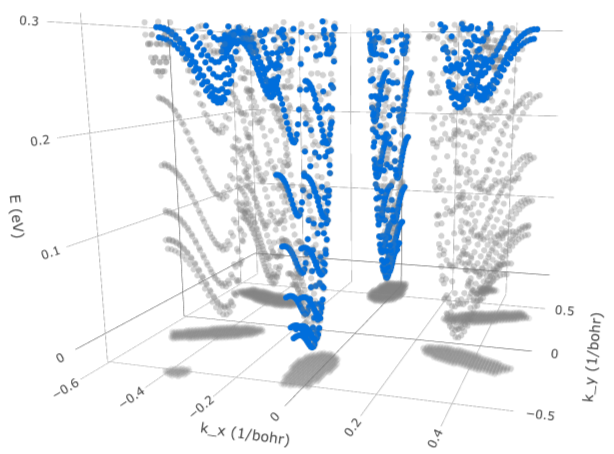}
\caption{Bands of strained (blue) and equilibrium (grey) blue phosphorene. The strain is $2\%$ in the $x$ direction. The shift in energy is $\approx 0.22$ eV. Here, the zero of the energy scale refers to the bottom of the lowest valley in each case. Like for antimonene, the valley at the K point is relatively low but shifts with strain. At equilibrium, another difference with arsenene is the shape of the six degenerate valleys, more elongated towards the M point. }
\label{fig:P2_bands}
\end{figure*}

\begin{figure*}[ht]
\includegraphics[width=0.85\textwidth]{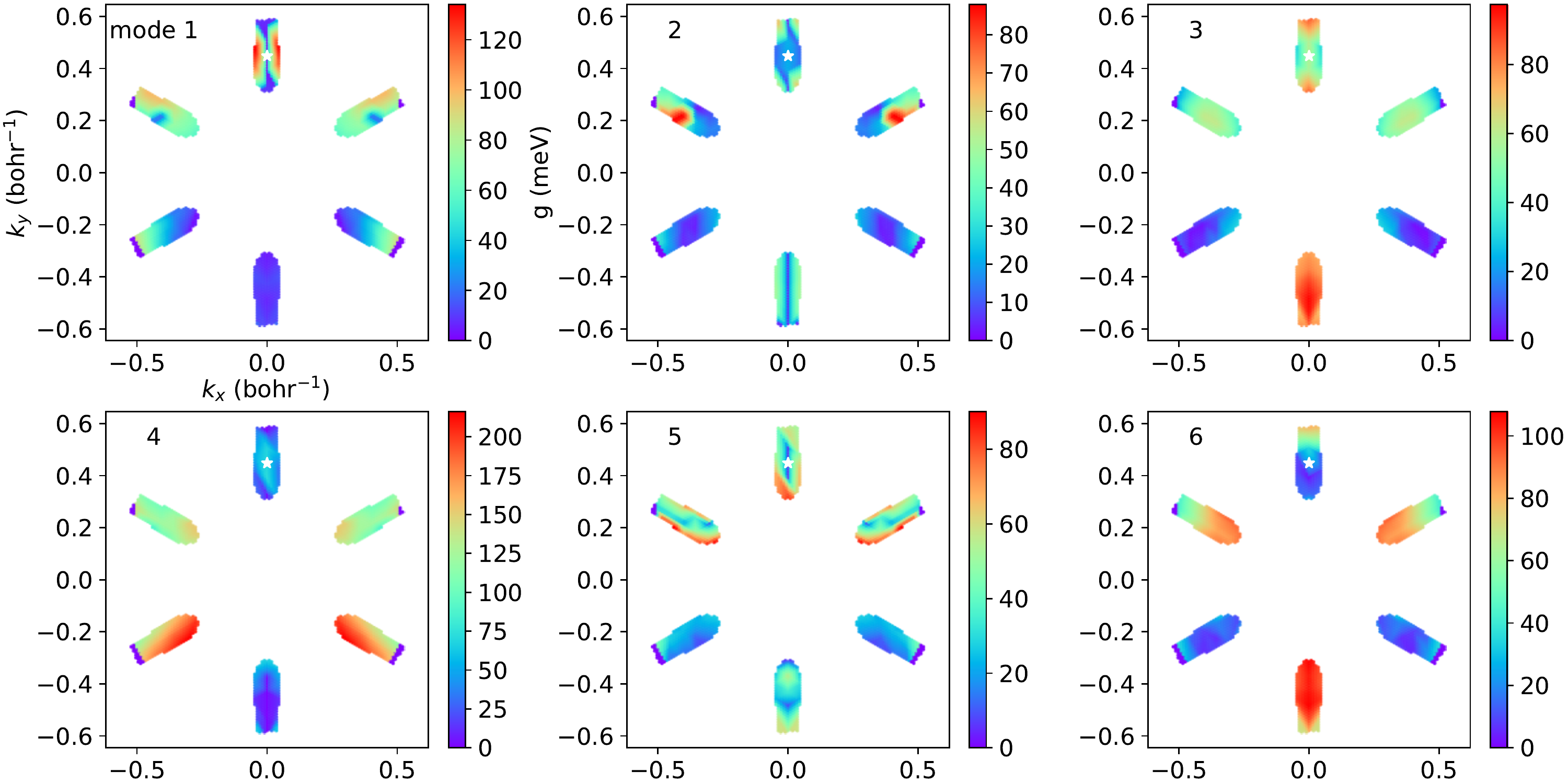}
\caption{Interpolated EPCs in blue phosphorene at equilibrium. The initial state $\bok$ considered here is indicated by a white star; the other points are the
possible final states, where the color of the point indicates the strength of the electron-phonon coupling matrix element (note the different color scales in each panel). The index of the phonon mode indicated at the top of each panel refers to a purely energetic ordering of the phonon modes associated with each transition. }
\label{fig:P2_EPC}
\end{figure*}

Fig. \ref{fig:P2_bands} shows the effect of strain on the bands of blue phosphorene. The shift in energy due to strain is significant ($+0.22$ eV). Fig. \ref{fig:P2_EPC} shows the electron-phonon couplings in blue phosphorene at equilibrium. We observe a large intravalley coupling for the first mode, which would decrease the relative importance of the intervalley modes and thus the enhancement.
The transport properties computed from those results are better than those of arsenene, with a room-temperature mobility of 150 cm$^2$V$^{-1}$s$^{-1}$ for the same doping of $n = 5/3 \times 10^{13}$ cm$^{-2}$. This is probably due to the more elongated nature of the valley mentioned in Fig. \ref{fig:P2_bands}. Overall, with less enhancement but better performance at equilibrium, one can expect similar performances for strained blue phosphorene.

\subsection{Raman spectrum of Arsenene}

\begin{figure*}[ht]
\includegraphics[width=0.47\textwidth]{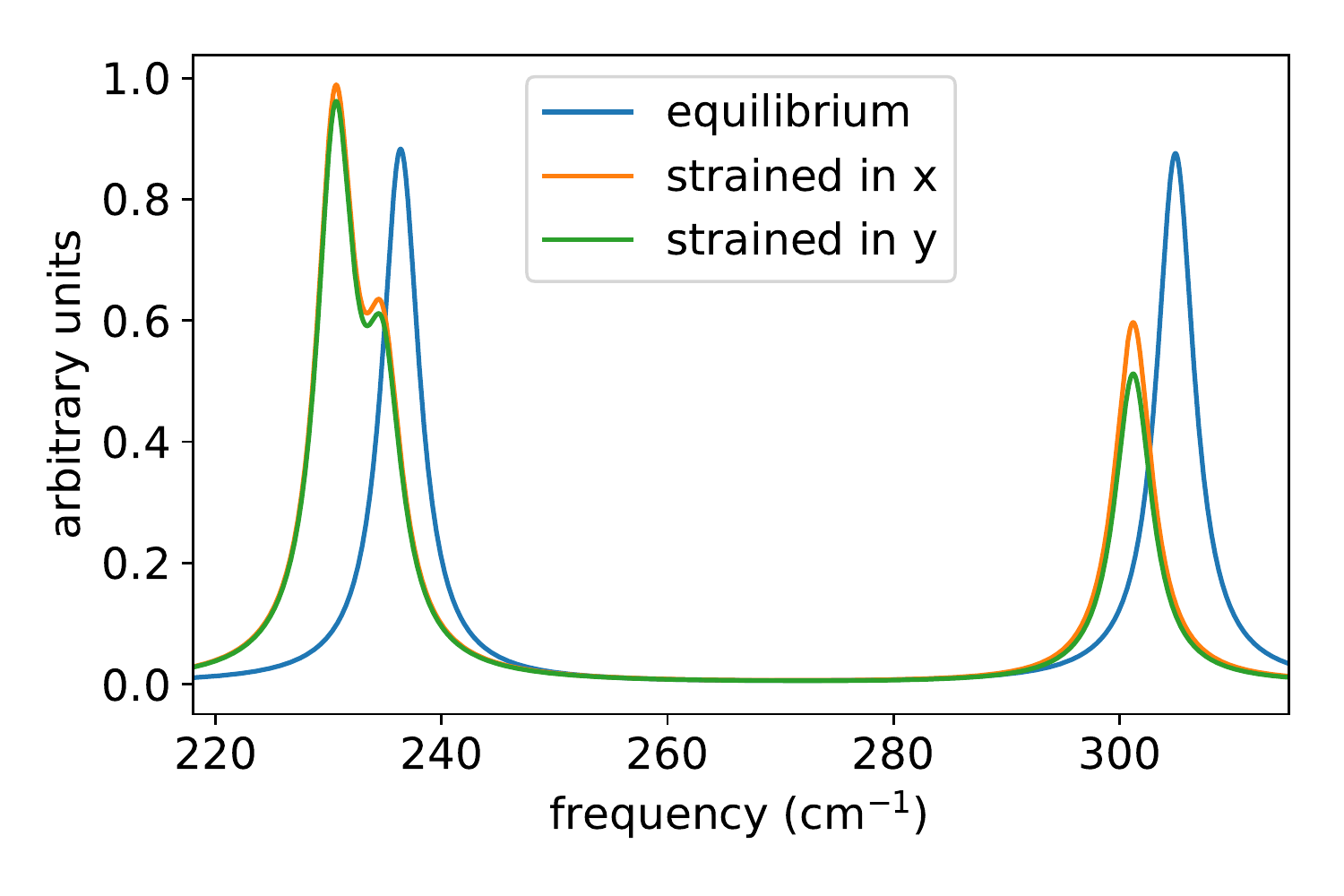}
\caption{Raman spectra for uniaxially strained ($2\%$) and equilibrium (unstrained) arsenene in the case of unpolarized light in back-scattering geometry. A Lorentzian function with a width of 2 cm$^{-1}$ has been used to approximate the $\delta$ function in Eq.~\eqref{eq:ramancross} and mimic the experimental linewidth.} \label{fig:raman_As}
\end{figure*}

\begin{figure*}[ht]
\includegraphics[width=0.47\textwidth]{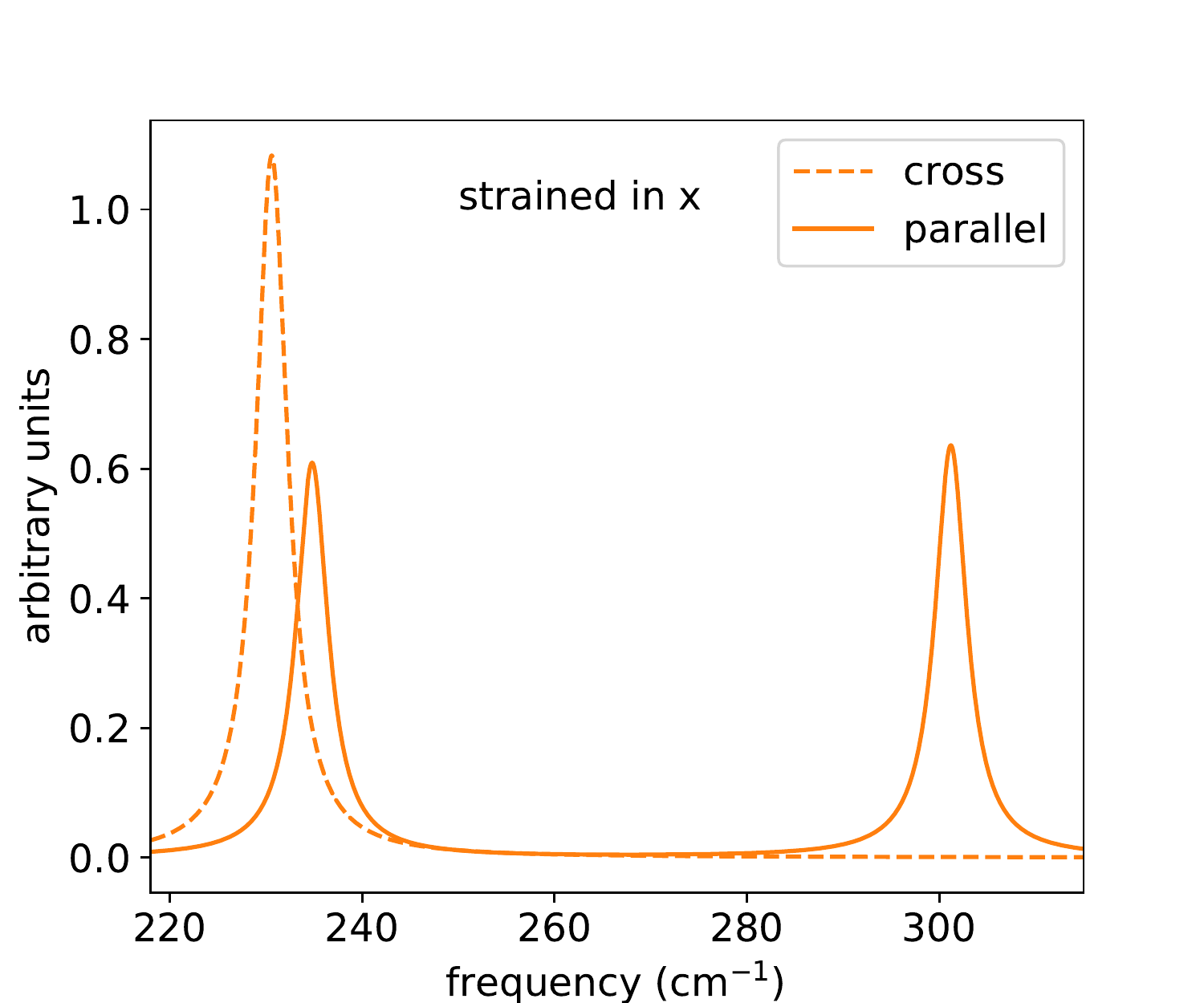}
\includegraphics[width=0.47\textwidth]{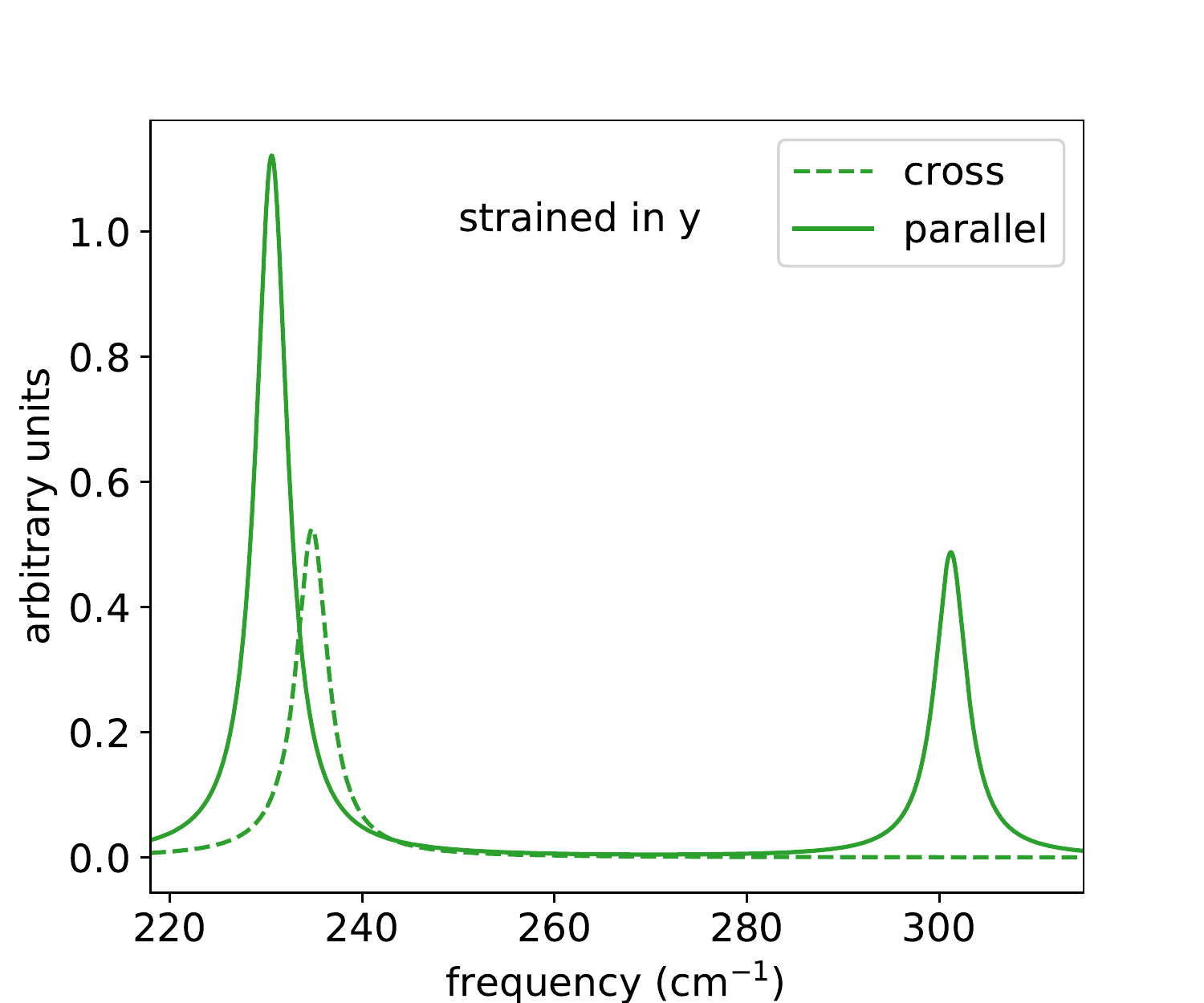}
\caption{Raman spectra for arsenene uniaxially strained ($2\%$) in the $x$ and $y$ direction for incident light polarized in the strain direction and scattered light in parallel (solid lines) or perpendicular (cross, dashed lines) directions.  } 
\label{fig:raman_As2}
\end{figure*}

In Table~\ref{tab:raman} and Fig.~\ref{fig:raman_As}, we give the frequencies of the Raman active modes and the Raman spectra for arsenene at equilibrium and strained in both $x$ and $y$ directions. The calculations are based on ab-initio, finite differences evaluations of the dielectric tensor derivatives. To compute the spectrum, we considered a back-scattering geometry  in non-resonant conditions\cite{Sosso2011,Wang2016}, for which the differential cross section can be expressed as:    
\begin{align}\label{eq:ramancross}
\frac{\delta^{2} \sigma}{\delta \Omega \delta \omega} &= \sum_{i} \frac{\omega_{S}^{4}}{c^{4}} \mid \textbf{e}_{S} \cdot R^{i} \cdot \textbf{e}_{I} \mid^{2} [n_{B}(\hbar \omega/k_{b}T) +1] \delta(\omega-\omega_{i})
\end{align}
where $\omega_{S}$ and $\omega_{i}$ are the frequencies of the scattered photons and of the $i$-th phonon respectively, $\textbf{e}_{S}$ and $\textbf{e}_{I}$ the polarization vectors of the scattered and incident light, and $R^{i}$ the Raman tensor. In the case of unpolarized light averaged over the in-plane directions the Raman cross section can be obtained from the equivalence:
\begin{align}
\mid \textbf{e}_{S} \cdot R^{i} \cdot \textbf{e}_{I} \mid^{2} &= \left(R^2_{xx}+R^2_{yy}+2R^2_{xy}\right)/2.
\end{align}
The spectra obtained under these conditions for  the strained and unstrained cases are reported in Fig.~\ref{fig:raman_As}. The magnitude of the shifts in frequency of the peaks should allow to follow the strain in the material. The order of the A$_g$ and B$_g$ modes in the strained case could be used to determine whether arsenene is strained in the $x$ (zig-zag) or $y$ (armchair) direction~\cite{Mohiuddin2009,Huang2009}. A$_g$ and B$_g$ modes can be distinguished using polarized incident light along the strain direction and measuring the scattered light polarized in the parallel or perpendicular (cross) direction. The Raman cross section in the two cases is given by:
\begin{align}
\mid \textbf{e}_{S} \cdot R^{i} \cdot \textbf{e}_{I} \mid^{2} &= \left(R_{xx}\cos^{2}(\theta)+R_{yy}\sin^{2}(\theta)+2R_{xy}\sin(2\theta)\right)^{2}
\end{align}
and
\begin{align}
\mid \textbf{e}_{S} \cdot R^{i} \cdot \textbf{e}_{I} \mid^{2} &= \left(R_{xy}\cos(2\theta)- \frac{R_{xx}-R_{yy}}{2}\sin(2\theta)\right)^{2}
\end{align}
where $\theta=0,\pi/2$ indicates the strain direction. 
The spectra for the two strain directions and polarization combinations are reported in Fig.~\ref{fig:raman_As2}.

\begin{table}[h]
\caption{Raman active modes of strained and unstrained arsenene.} 
\begin{tabular}{ c c c }
\hline
Material & Frequency (cm$^{-1}$) & Character  \\ 
\hline
As equilibirum & 236.4 & E$_g$ \\
As equilibrium & 304.9 & A$_{1g}$ \\
\hline
As 2\% strain $x$ & 230.6 & B$_g$ \\
As 2\% strain $x$ & 234.8 & A$_g$ \\
As 2\% strain $x$ & 301.2 & A$_g$ \\
\hline
As 2\% strain $y$ & 230.5 & A$_g$ \\
As 2\% strain $y$ & 234.9 & B$_g$ \\
As 2\% strain $y$ & 301.2 & A$_g$ \
\end{tabular}
\label{tab:raman}
\end{table}

\clearpage

\bibliographystyle{myapsrev}
\bibliography{As}

\end{document}